\input{aipcheck}

\documentclass[
    ,final            
  ]
  {aipproc}

\layoutstyle{8x11single}

\usepackage{amsmath}
\begin{document}

\title{The Gravitational Field of a Twisted Skyrmion String}

\classification{00.04, 10.12}
\keywords      {twisted Skyrmion string, gravitational field, the Einstein field equations}

\author{Miftachul Hadi$^{e}$}{
  address={Department of Mathematics, Universiti Brunei Darussalam, Jalan Tungku Link BE1410, Gadong, Negara Brunei Darussalam}
,altaddress={Physics Research Centre, Indonesian Insitute of Sciences (LIPI), Puspiptek,
Serpong, Tangerang Selatan 15314, Banten, Indonesia}
,altaddress={Department of Physics, School of Natural Sciences, Ulsan National Institute of
Science and Technology (UNIST), 50, UNIST-gil, Eonyang-eup, Ulju-gun, Ulsan, South Korea}
,altaddress={Institute of Modern Physics, Chinese Academy of Sciences, 509 Nanchang Rd.,
Lanzhou 730000, China}
}

\author{Malcolm Anderson}{
  address={Department of Mathematics, Universiti Brunei Darussalam, Jalan Tungku Link
BE1410, Gadong, Negara Brunei Darussalam}
}

\author{Andri Husein}{address={Department of Physics, University of Sebelas Maret, Jalan Ir. Sutami 36 A,
Surakarta 57126, Indonesia} 
{$^{e}$E-mail: itpm.id@gmail.com} 
}

\begin{abstract}
In this paper we study the gravitational field of a straight string generated from a class of nonlinear sigma models, specifically the Skyrme model with a twist (the twisted Skyrmion string). The twist term, $mkz$, is included to stabilize the vortex solution. To model the effects of gravity, we replace the Minkowski tensor, $\eta^{\mu\nu}$, in the standard Skyrme Lagrangian density with a space-time metric tensor, $g^{\mu\nu}$, assumed to be static and cylindrically symmetric. The Einstein equations for the metric and field components are then derived. This work is still in progress.
\end{abstract}

\maketitle


\section{A TWISTED SKYRME LAGRANGIAN}
We are interested in constructing the space-time generated by a twisted Skyrmion string. Without gravity, the standard Skyrme Lagrangian density of the system is 
\begin{eqnarray}\label{1}
\mathcal{L}_1
&=& \frac{1}{2\lambda^2}\eta^{\mu\nu}\partial_\mu\phi.\partial_\nu\phi -K_s~\eta^{\kappa\lambda}\eta^{\mu\nu}(\partial_\kappa\phi\times\partial_\mu\phi).(\partial_\lambda\times\partial_\nu\phi)
\end{eqnarray}
where the second term on the right hand side of (\ref{1}) is the Skyrme term. 

To add gravity, we replace $\eta^{\mu\nu}$ in $\mathcal{L}_1$ with a space-time metric tensor, $g^{\mu\nu}$, which in view of the time-independence and cylindrical symmetry of the assumed vortex solution is taken to be a function of $r$ alone. Metric tensor, $g^{\mu\nu}$, is of course the inverse of the covariant metric tensor, $g_{\mu\nu}$, of the space-time where $g^{\mu\nu}=(g_{\mu\nu})^{-1}$. We use a cylindrical coordinate system $(t,r,\theta,z)$, where $t$ and $z$ have unbounded range, $r\in[0,\infty)$ and $\theta\in[0,2\pi)$. 
The components of the metric tensor
\begin{eqnarray}\label{2}
g_{\mu\nu}
=
\begin{pmatrix}
g_{tt} & 0      & 0                & 0  \\
0      & g_{rr} & 0                & 0  \\
0      & 0      & g_{\theta\theta} & g_{\theta z} \\
0      & 0     & g_{z\theta}      & g_{zz}
\end{pmatrix}
\end{eqnarray}
are all functions of $r$, and the presence of the off-diagonal components $g_{\theta z}=g_{z\theta}$ reflects the twist in the space-time.

The Lagrangian we will be using is
\begin{eqnarray}\label{3}
\mathcal{L}_2
&=& \frac{1}{2\lambda^2}~(g^{\mu\nu}~\partial_\mu f~\partial_\nu f+\sin^2f~g^{\mu\nu}~\partial_\mu g~\partial_\nu g) \nonumber\\
&&+~2K_s~\sin^2f~[(g^{\mu\nu}~\partial_\mu f~\partial_\nu f)(g^{\kappa\lambda}~\partial_\kappa g~\partial_\lambda g) -2\sin^2f~(g^{\mu\nu}~\partial_\mu f~\partial_\nu g)^2]
\end{eqnarray}
with $f=f(r)$ and $g=n\theta+mkz$. Here, $mkz$ is twist term. 

\section{THE EINSTEIN EQUATION}
We need to solve
\begin{itemize}
\item[(i)] the Einstein equations
\begin{eqnarray}\label{4}
G_{\mu\nu}
&=& -\frac{8\pi G}{c^4}~T_{\mu\nu}  
\end{eqnarray}
where the stress-energy tensor of the vortex, $T_{\mu\nu}$, is defined by
\begin{eqnarray}\label{5}
T_{\mu\nu}
&\equiv&2\frac{\partial\mathcal{L}_2}{\partial g^{\mu\nu}}-g_{\mu\nu}~\mathcal{L}_2
\end{eqnarray}
and
\begin{eqnarray}\label{6}
G_{\mu\nu} 
&=& R_{\mu\nu}-\frac{1}{2}g_{\mu\nu}~R 
\end{eqnarray}
with
$R^{\mu\nu}$ the Ricci tensor and 
\begin{eqnarray}\label{7}
R=g_{\mu\nu}~R^{\mu\nu} = g^{\mu\nu}~R_{\mu\nu}
\end{eqnarray}
the Ricci scalar; and
\item[(ii)] the field equations for $f$ and $g$
\begin{eqnarray}\label{8}
\nabla^\mu\frac{\partial\mathcal{L}_2}{\partial(\partial f/\partial x^\mu)}=\frac{\partial\mathcal{L}_2}{\partial f}~~~\text{and}~~~\nabla^\mu\frac{\partial\mathcal{L}_2}{\partial(\partial g/\partial x^\mu)}=\frac{\partial\mathcal{L}_2}{\partial g}
\end{eqnarray}
\end{itemize}
However, the field equations for $f$ and $g$ are in fact redundant, as they are satisfied identically whenever the Einstein equations are satisfied, by virtue of the Bianchi identities $\nabla_\mu G^{\mu}_{\nu}=0$. So only the Einstein equations will be considered in this section.

To simplify the Einstein equations, we first choose a gauge condition that narrows down the form of the metric tensor. The gauge condition preferred here is that
\begin{eqnarray}\label{10}
g_{\theta\theta}~g_{zz}-(g_{\theta z})^2=r^2
\end{eqnarray}
The geometric significance of this choice is that the determinant of the 2-metric tensor projected onto the surfaces of constant $t$ and $z$ is $r^2$, and so the area element on these surfaces is just $r~dr~d\theta$.

As a further simplification, we write
\begin{eqnarray}\label{11}
g_{tt}=A^2;~~~~~g_{rr}=-B^2;~~~~~g_{\theta\theta}=-C^2;~~~~~g_{\theta z}=\omega
\end{eqnarray}
and so
\begin{eqnarray}\label{13}
g_{zz}=-\left(\frac{r^2+\omega^2}{C^2}\right).
\end{eqnarray}
The metric tensor, $g_{\mu\nu}$, therefore has the form
\begin{eqnarray}\label{14}
g_{\mu\nu}=
\begin{pmatrix}
A^2  &  0     &    0     &  0  \\
0    &  -B^2  &    0     &  0  \\
0    &  0     &  -C^2    &  \omega  \\
0    &  0     &   \omega & -\left(\frac{r^2+\omega^2}{C^2}\right)
\end{pmatrix}
\end{eqnarray}
Substituting (\ref{14}) into $\mathcal{L}_2$ gives
\begin{eqnarray}\label{15}
\mathcal{L}_2
&=&-~\frac{1}{2\lambda^2}\left\{B^{-2}f'^2+\sin^2f[n^2(1+\omega^2/r^2)C^{-2} +2mkn~r^{-2}\omega+m^2k^2r^{-2}C^2]\right\}\nonumber\\
&&+~2K_s~\sin^2f\left\{B^{-2}f'^2[n^2(1+\omega^2/r^2)C^{-2} +2mkn~r^{-2}\omega+m^2k^2r^{-2}C^2]\right\} 
\end{eqnarray}
From equations (\ref{5}), (\ref{14}) and (\ref{15}), the non-zero components of the stress-energy tensor, $T_{\mu\nu}$, are:
\begin{eqnarray}\label{16}
T_{tt}
=\frac{1}{2\lambda^2}~A^2B^{-2}f'^2  -A^2\left( -\frac{1}{2\lambda^2} +2K_s B^{-2}f'^2\right)\sin^2f[n^2(1+\omega^2r^{-2})C^{-2} +2mkn~\omega r^{-2} +(mk)^2C^2 r^{-2}] 
\end{eqnarray}
\begin{eqnarray}\label{17}
T_{rr}
=\frac{1}{2\lambda^2}~f'^2  -B^2\left(\frac{1}{2\lambda^2}  +2K_sB^{-2}f'^2\right)\sin^2f[n^2(1+\omega^2r^{-2})C^{-2} +2mkn~\omega r^{-2} +(mk)^2C^2r^{-2}] 
\end{eqnarray}
\begin{eqnarray}\label{18}
T_{\theta\theta}
=-~\frac{1}{2\lambda^2}~ B^{-2}C^2f'^2 +\left(-\frac{1}{2\lambda^2} +2K_sB^{-2}f'^2\right)\sin^2f[n^2(-1 +\omega^2r^{-2}) +2mkn~\omega C^2 r^{-2}  +(mk)^2r^{-2}C^4] 
\end{eqnarray}
\begin{eqnarray}\label{19}
T_{\theta z}
=\frac{1}{2\lambda^2}~B^{-2}\omega f'^2+\left(\frac{1}{2\lambda^2} -2K_s B^{-2}f'^2\right)\sin^2f[n^2(1 +\omega^2r^{-2})\omega C^{-2}  +2mkn(1 +\omega^2r^{-2}) +(mk)^2r^{-2}\omega C^2]
\end{eqnarray}
and
\begin{eqnarray}\label{20}
T_{zz}
&=&-~\frac{1}{2\lambda^2}~(\omega^2 +r^2)B^{-2}C^{-2}f'^2 +\left(-\frac{1}{2\lambda^2} +2K_s B^{-2}f'^2\right)\sin^2f\nonumber\\
&&\times~[n^2r^2 (1 +\omega^2r^{-2})^2C^{-4} +2mkn(1 +r^{-2}\omega^2)\omega C^{-2} +(mk)^2(-1 +r^{-2}\omega^2)] 
\end{eqnarray}
where $f'=df/dr$. Note that $T_{\theta z}$ is in general non-zero, provided that either $mk$ or $\omega$ is non-zero. In fact, $mk$ acts as a source term for $\omega$, as will be seen in equation (\ref{36}) below. The twist in the vortex is therefore solely responsible for a non-zero circular stress $T_{\theta z}$.

In component form, the Einstein equations (\ref{4}) read
\begin{eqnarray}\label{21}
G_{tt}
&=&-\frac{8\pi G}{c^4}~T_{tt};~~~G_{rr}= -\frac{8\pi G}{c^4}~T_{rr};~~~G_{\theta\theta}=-\frac{8\pi G}{c^4}~T_{\theta\theta};~~~G_{\theta z}=-\frac{8\pi G}{c^4}~T_{\theta z}
\end{eqnarray}
and
\begin{eqnarray}\label{23}
G_{zz}
&=&-\frac{8\pi G}{c^4}~T_{zz}
\end{eqnarray}
The non-zero components of the Einstein tensor, $G_{\mu\nu}$, corresponding to the metric tensor (\ref{6}) are
\begin{eqnarray}\label{24}
G_{tt}
&=&R_{tt} -\frac{1}{2}g_{tt}~R = -~\frac{A^2B'}{rB^3}-\frac{A^2C'}{rB^2C}+\frac{A^2C'^2}{B^2C^2}(1+r^{-2}\omega^2) -\frac{A^2\omega\omega'C'}{r^2B^2C}+\frac{A^2\omega'^2}{4r^2B^2}
\end{eqnarray}
\begin{eqnarray}\label{25}
G_{rr}
&=&R_{rr} -\frac{1}{2}g_{rr}~R = -\frac{A'}{rA}-\frac{C'}{rC}+\frac{C'^2}{C^2}(1+r^{-2}\omega^2) -\frac{\omega\omega'C'}{r^2C} +\frac{\omega'^2}{4r^2}
\end{eqnarray}
\begin{eqnarray}\label{26}
G_{\theta\theta}
&=&R_{\theta\theta} -\frac{1}{2}g_{\theta\theta}~R =-~\frac{C^2A'}{rAB^2}-\frac{C^2A''}{AB^2}+\frac{C^2B'}{rB^3}+\frac{C^2A'B'}{AB^3} +\frac{2CC'}{rB^2}+\frac{CC''}{B^2} -\frac{C'^2}{B^2}(2+3r^{-2}\omega^2) \nonumber\\
&&+~\frac{CA'C'}{AB^2} -\frac{CB'C'}{B^3}+\frac{3\omega\omega'CC'}{r^2B^2} -\frac{3C^2\omega'^2}{4r^2B^2}
\end{eqnarray}
\begin{eqnarray}\label{27}
G_{\theta z}
&=&R_{\theta z} -\frac{1}{2}g_{\theta z}~R =\frac{\omega A'}{rAB^2} -\frac{\omega'A'}{2AB^2} +\frac{\omega A''}{AB^2} -\frac{\omega B'}{rB^3} +\frac{\omega'B'}{2B^3} -\frac{\omega A'B'}{AB^3} -\frac{3\omega C'}{rB^2C} -\frac{3\omega^2\omega'C'}{r^2B^2C} \nonumber\\
&&+~\frac{3\omega C'^2}{B^2C^2}(1+r^{-2}\omega^2)+\frac{\omega'}{2rB^2}-\frac{\omega''}{2B^2} +\frac{3\omega\omega'^2}{4r^2B^2}
\end{eqnarray}
with $G_{\theta z}=G_{z\theta}$, and
\begin{eqnarray}\label{28}
G_{zz}
&=&R_{zz} -\frac{1}{2}g_{zz}~R =-\frac{\omega^2A'}{rAB^2C^2} -\frac{r^2A''}{AB^2C^2}(1+r^{-2}\omega^2) +\frac{r^2A'B'}{AB^3C^2}(1+r^{-2}\omega^2) -\frac{r^2A'C'}{AB^2C^3}(1+r^{-2}\omega^2) \nonumber\\
&&+~\frac{\omega^2B'}{rB^3C^2} +\frac{r^2B'C'}{B^3C^3}(1+r^{-2}\omega^2) +\frac{4\omega^2 C'}{rB^2C^3} -\frac{r^2C''}{B^2C^3}(1+r^{-2}\omega^2) -\frac{3\omega^2C'^2}{B^2C^4}(1+r^{-2}\omega^2) \nonumber\\
&&+~\frac{\omega\omega'A'}{AB^2C^2} -\frac{\omega\omega'B'}{B^3C^2} -\frac{\omega\omega'C'}{B^2C^3}(1-3r^{-2}\omega^2) -\frac{\omega\omega'}{rB^2C^2} +\frac{\omega\omega''}{B^2C^2} +\frac{\omega'^2}{4B^2C^2}(1-3r^{-2}\omega^2) 
\end{eqnarray}
Substituting (\ref{24})-(\ref{28}) and (\ref{16})-(\ref{20}) into equations (\ref{21})-(\ref{23}) now gives  
\begin{eqnarray}\label{29}
-\frac{8\pi G}{c^4}~T_{tt} 
&=&-~\frac{A^2B'}{rB^3}-\frac{A^2C'}{rB^2C}+\frac{A^2C'^2}{B^2C^2}(1+r^{-2}\omega^2) -\frac{A^2\omega\omega'C'}{r^2B^2C}+\frac{A^2\omega'^2}{4r^2B^2} 
\end{eqnarray}
\begin{eqnarray}\label{30}
-\frac{8\pi G}{c^4}~T_{rr} 
&=&-~\frac{A'}{rA}-\frac{C'}{rC}+\frac{C'^2}{C^2}(1+r^{-2}\omega^2) -\frac{\omega\omega'C'}{r^2C} +\frac{\omega'^2}{4r^2} 
\end{eqnarray}
\begin{eqnarray}\label{31}
-\frac{8\pi G}{c^4}~T_{\theta\theta} 
&=&-\frac{C^2A'}{rAB^2}-\frac{C^2A''}{AB^2}+\frac{C^2B'}{rB^3}+\frac{C^2A'B'}{AB^3} +\frac{2CC'}{rB^2}+\frac{CC''}{B^2} -\frac{C'^2}{B^2}(2+3r^{-2}\omega^2) +\frac{CA'C'}{AB^2} -\frac{CB'C'}{B^3}  \nonumber\\
&&+~\frac{3\omega\omega'CC'}{r^2B^2} -\frac{3C^2\omega'^2}{4r^2B^2} 
\end{eqnarray}
\begin{eqnarray}\label{32}
-\frac{8\pi G}{c^4}~T_{\theta z}
&=&\frac{\omega A'}{rAB^2} -\frac{\omega'A'}{2AB^2} +\frac{\omega A''}{AB^2} -\frac{\omega B'}{rB^3} +\frac{\omega'B'}{2B^3} -\frac{\omega A'B'}{AB^3} -\frac{3\omega C'}{rB^2C} -\frac{3\omega^2\omega'C'}{r^2B^2C} \nonumber\\
&&+~\frac{3\omega C'^2}{B^2C^2}(1+r^{-2}\omega^2)+\frac{\omega'}{2rB^2}-\frac{\omega''}{2B^2} +\frac{3\omega\omega'^2}{4r^2B^2}  
\end{eqnarray}
and
\begin{eqnarray}\label{33}
-\frac{8\pi G}{c^4}~T_{zz} 
&=&-\frac{\omega^2A'}{rAB^2C^2} -\frac{r^2A''}{AB^2C^2}(1+r^{-2}\omega^2) +\frac{r^2A'B'}{AB^3C^2}(1+r^{-2}\omega^2) -\frac{r^2A'C'}{AB^2C^3}(1+r^{-2}\omega^2) \nonumber\\
&&+~\frac{\omega^2B'}{rB^3C^2} +\frac{r^2B'C'}{B^3C^3}(1+r^{-2}\omega^2) +\frac{4\omega^2 C'}{rB^2C^3} -\frac{r^2C''}{B^2C^3}(1+r^{-2}\omega^2) -\frac{3\omega^2C'^2}{B^2C^4}(1+r^{-2}\omega^2) +\frac{\omega\omega'A'}{AB^2C^2} \nonumber\\
&&-~\frac{\omega\omega'B'}{B^3C^2}  -\frac{\omega\omega'C'}{B^2C^3}(1-3r^{-2}\omega^2) -\frac{\omega\omega'}{rB^2C^2} +\frac{\omega\omega''}{B^2C^2} +\frac{\omega'^2}{4B^2C^2}(1-3r^{-2}\omega^2) 
\end{eqnarray}

These equations can be rearranged as source equations for the second derivatives of $A$, $C$ and $\omega$ and the first derivatives of $B$ and $f$ as follows:
\begin{eqnarray}\label{34}
A''
&=&-~\frac{A'}{2r}  +\frac{A'B'}{B}  +\frac{AB'}{2rB}  +\frac{AC'}{rC}  +\frac{\omega\omega'AC'}{r^2C}  -\frac{AC'^2}{C^2}(1 +r^{-2}\omega^2 )  -\frac{A\omega'^2}{4r^2}  -\frac{\kappa}{2\lambda^2}Af'^2 
\end{eqnarray}
\begin{eqnarray}\label{35}
C''
&=&-~\frac{C'}{r}  +\frac{CA'}{2rA}  -\frac{A'C'}{A}  -\frac{CB'}{2rB}  +\frac{B'C'}{B}  -\frac{2\omega\omega'C'}{r^2}  +\frac{C'^2}{C}(1 +2r^{-2}\omega^2)  +\frac{C\omega'^2}{2r^2} \nonumber\\
&&-~\kappa B^2 C^{-1}r^{-2}[(n\omega +C^2km)^2 -(nr)^2]\left(-\frac{1}{2\lambda^2}  +2K_sB^{-2}f'^2\right)\sin^2f 
\end{eqnarray}
\begin{eqnarray}\label{36}
\omega''
&=& \frac{\omega'}{r}  +\frac{\omega\omega'^2}{r^2}  +\frac{\omega A'}{rA}  -\frac{\omega'A'}{A}  -\frac{\omega B'}{rB}  +\frac{\omega'B'}{B}-\frac{4\omega C'}{rC}  -\frac{4\omega^2\omega'C'}{r^2C}  +\frac{4\omega C'^2}{C^2}(1 +r^{-2}\omega^2)\nonumber\\
&&-~2\kappa B^2 C^{-2}r^{-2}\left\{\omega[(n\omega +C^2km)^2 +(nr)^2] +2C^2kmn~r^2\right\}\left(-\frac{1}{2\lambda^2}  +2K_sB^{-2}f'^2\right)\sin^2f 
\end{eqnarray}
\begin{eqnarray}\label{37}
B'
&=& A^{-1}BA'  +\frac{\kappa}{\lambda^2}B^3C^{-2}r^{-1}[(n\omega +C^2km)^2  +(nr)^2]\sin^2f
\end{eqnarray}
and
\begin{eqnarray}\label{38}
f'^2
&=& \left\{\kappa^{-1}\left[\frac{A'}{rA}  +\frac{C'}{rC}  -\frac{C'^2}{C^2}(1 +r^{-2}\omega^2)  +\frac{\omega\omega'C'}{r^2C}  -\frac{\omega'^2}{4r^2}\right]  +\frac{1}{2\lambda^2}B^2C^{-2}r^{-2}[(n\omega +C^2km)^2 +(n\omega)^2]\sin^2f\right\}\nonumber\\
&&\times~\left\{\frac{1}{2\lambda^2}  -2K_sC^{-2}r^{-2}[(n\omega  +C^2km)^2  +(nr)^2]\sin^2f\right\}^{-1}
\end{eqnarray}
where $\kappa=8\pi G/c^4$.

\section{SOLUTION OF THE EINSTEIN FIELD EQUATIONS}
In order to solve equations, the Einstein field equations (\ref{34})-(\ref{38}), we require boundary conditions on the functions $A$, $B$, $C$, $\omega$ and $f$ for small $r$ and in the limit as $r$ tends to $\infty$. In particular, $\omega$ should vanish both at $r=0$ (at least as rapidly as $r^2$, so as to preserve elementary flatness on the axis of symmetry) and in the limit as $r\to\infty$ (to allow for a locally Minkowski vacuum at space-like infinity). It is evident from equation (\ref{36}) that $\omega$ will be zero everywhere unless $mk\neq0$, and so the vortex twist acts as a source for the space-time twist as mentioned above. This work is still in progress.

\begin{theacknowledgments}
We thank the Editor and the referee for their comments. This research is funded fully by Graduate Research Scholarship Universiti Brunei Darussalam (GRS UBD). This support is greatly appreciated.
\end{theacknowledgments}



\bibliographystyle{aipproc}   


\IfFileExists{\jobname.bbl}{}
 {\typeout{}
  \typeout{******************************************}
  \typeout{** Please run "bibtex \jobname" to optain}
  \typeout{** the bibliography and then re-run LaTeX}
  \typeout{** twice to fix the references!}
  \typeout{******************************************}
  \typeout{}
 }

\end{document}